\documentclass[twocolumn,showpacs,preprintnumbers,amsmath,amssymb]{revtex4}
\usepackage{graphicx}
\usepackage{dcolumn}
\usepackage{bm}
\usepackage{marginnote}

\begin{document}

\preprint{xxxx}

\title{Interaction of Ultra Relativistic $e^{-} e^{+}$ Fireball Beam with Plasma}

\author{P. Muggli$^{1}$}
\email{muggli@usc.edu}
\author{S. F. Martins$^{2}$}%
\author{N. Shukla$^{2}$}%
\author{J. Vieira$^{2}$}%
\author{L. O. Silva$^{2}$}%
\email{luis.silva@ist.utl.pt}
\affiliation{$^{1}$University of Southern California, Los Angeles, CA 90089, USA}%
\affiliation{$^{2}$GoLP/Instituto de Plasmas e Fus\~ao Nuclear, Instituto Superior T\'ecnico, Lisbon, Portugal}%

\date{\today}


\begin{abstract}
{\it Ab initio} simulations of the propagation in a plasma of a soon to be available relativistic electron-positron beam or {\it fireball beam} provide an effective mean for the study of microphysics relevant to astrophysical scenarios. We show that the current filamentation instability associated with some of these scenarios reaches saturation after only 10~cm of propagation in a typical laboratory plasma with a density  $\sim 10^{17}$~cm$^{-3}$. The different regimes of the instability, from the purely transverse to the mixed mode filamentation, can be accessed by varying the background plasma density. The instability generates large local plasma gradients, intense transverse magnetic fields, and enhanced emission of radiation. We suggest that these effects may be observed experimentally for the first time.
 
\end{abstract}

\maketitle

Several astrophysical scenarios lead to extreme physical regimes, typically observed on Earth in the form of radiation and cosmic rays. These regimes encompass a set of phenomena such as magnetic field generation, shock formation, energy transfer processes, and non-thermal particle acceleration (for a review, see  \cite{jones91}). In the particular case of the fireball model of gamma ray bursts (GRBs) \cite{piran04mesz93}, the kinetic energy of an ultra-relativistic plasma shell, with an arbitrary mixture of electrons-positrons-ions ($e^- e^+ p^+$),  is converted into radiation as moving shells collide, but the specific conversion mechanism is still an open question. Relativistic flows are also frequent in shock waves and pulsar wind nebulae, where relativistic shells interact with a background plasma \cite{kazimura98}. 
It is very difficult to reproduce these astrophysical conditions in the laboratory, and the studies of the nonlinear physical phenomena are essentially simplified analytical models and numerical simulations (see \cite{ellison04keshet06, spitk08, martins09} and references therein). Identifying the laboratory conditions that can validate the conclusions reached in previous studies is thus of paramount importance. Progress in laser technology, for instance, already suggests the possibility to explore experimentally scaled-down astrophysical phenomena in laser-plasma interactions \cite{woolsey01bulanov08chen09}.

In this Letter we focus on a scenario similar to that widely believed to be present in, and at the origin of GRBs, by examining the collision of a relativistic $e^-e^+$ beam or neutral plasma (that we call a {\em fireball beam}) mimicking a realistic plasma shell, with a static plasma consisting of $e^-$ and $p^+$. The interaction leads to current filamentation instability (CFI), or Weibel instability \cite{weibel59,medvedev99}, which generates very large magnetic fields as the beam plasma interaction evolves. The self-consistent evolution of electric (E) and magnetic (B) fields, and the resulting radiation generation as particles propagate in CFI driven turbulence are observed. Here, we consider conditions that will soon become available in the laboratory, and focus on the direct observation of the CFI of an $e^-e^+$ neutral beam. We also examine the possible experimental detection of the nonlinear stage of this instability as a function of its key parameters, through imaging of the associated plasma gradients, detection of the beam radiation, and the B-field of the beam filaments. Our results show that the role and the dynamics of the plasma microinstabilities associated with the collision of a relativistic fireball with a plasma may be probed in the laboratory.

Recent theoretical \cite{medvedev99, gruzinov99} and numerical results \cite{silva03, frederiksen04nishikawa05} have shown the relevance of the CFI for GRBs scenarios and for the onset of relativistic shocks in unmagnetized plasmas \cite{spitk08,martins09}. Moreover, the consequences of the excitation of the mixed mode, or tilted filamentation \cite{bretsilva}, in the long time evolution of the generated E/B-fields remains to be addressed; it was suggested that this preferential mode excitation will lead to significant beam spraying \cite{silva03}, but the impact on the saturated level of the fields was not discussed. Experimental evidence for the radiation signatures from Weibel turbulence \cite{medvedev05}, the energy transfer rate from the fireball to the B-field, and the long time evolution of the self-generated E and B-fields are critical inputs to the existing models, to perform direct comparisons with astronomical observations, and to assess the relevance of the CFI to relativistic astrophysics. 

Relativistic $e^-$ beams are available in many laboratories around the world, while $e^+$ beams are not. Recent numerical studies of a plasma-based, $e^-e^+$ accelerator concept based on the plasma wakefield accelerator (PWFA) \cite{chen85} have revealed that it may be advantageous to accelerate a $e^+$ bunch on the wake driven by an $e^-$ bunch \cite{Lotov07}. Ultra-relativistic $e^-$ and $e^+$ bunches suitable to test this acceleration scheme are available at the SLAC National Accelerator Laboratory. For this test, the distance between the $e^-$ and the $e^+$ bunch must be adjustable and on the order of the plasma wavelength or about $100~\mu$m. A double or \textit{sailboat} magnetic chicane has been developed \cite{decker}, that allows for the adjustment of the spacing between the two bunches, and may be used to overlap the two bunches with equal charge, both in space and time to effectively create a relativistic {\em fireball beam}. This will make possible the first ever collision between relativistic neutral plasmas in the laboratory: a relativistic $e^-e^+$ plasma onto an $e^-p^+$ plasma at rest, separating the effects of the space charge fields associated with a charged beam. 

We investigate the propagation of the SLAC {\em fireball beam} in a pre-formed plasma with numerical simulations performed with the fully relativistic, fully electromagnetic, and massivelly parallel particle-in-cell (PIC) code OSIRIS \cite{fonseca08}. This simulation framework has been extensively used for studies of laser/beam plasma interaction (e.g., \cite{dodd02}), and astrophysical regimes (e.g., \cite{silva03, silva06fonseca03}), among others. The system is studied numerically with a $205\times205\times82~\mu$m$^3$ window moving at the speed of light along the z-direction, and discretized in $400\times400\times80$ cells with absorbing boundary conditions for the fields and for the particles in the transverse x, y directions. The {\em fireball beam} is defined with Gaussian profiles in all directions with rms sizes: $\sigma_x = \sigma_y = 2\sigma_z = 2~c/\omega_{\mathrm{pe}} = 20.4~\mu$m, where $\omega_{\mathrm{pe}}=(n_{e}e^2/\epsilon_{0}m_{e})^{1/2}$ is the $e^-$ pulsation of the rest plasma with density $n_{e}=2.7\times10^{17}$~cm$^{-3}$. The standard beam used in the simulations has $1.8\times 10^{10}$ $e^-$, and the same number of $e^+$, all with an incoming energy of $29$~GeV and a normalized emittance of $2\times 10^{-5}$~m-rad, corresponding to a peak beam density $n_b = n_e$, 
and a transverse thermal spread $v_\mathrm{th}/c = 1.7 \times 10^{-5}$. A total of $\sim 7\times 10^7$ simulation particles (plasma and beam) is pushed for $\sim 10^4~c/\omega_{\mathrm{pe}} \simeq 10$~cm of pre-formed plasma (20~cm were also simulated to confirm the saturated state parameters). The time step is $0.033/\omega_{\mathrm{pe}}$. 
The neutrality of the beam guaranties its propagation at constant radius. A background of fixed $p^+$ is assumed for the pre-formed plasma: quantitative variations below 1\% were obtained for the standard case when using mobile $p^+$. Note also that, as in astrophysics, these parameters correspond to a collisionless fireball-plasma interaction: $\nu_{ei}/\omega_{\mathrm{pe}}=\mathcal O(10^{-18})$, where $\nu_{ei}$ is the beam $e^{-}$, $e^{+}$-background $p^+$ collision frequency. 

\begin{figure}[htp]
\centering
\includegraphics[scale=0.35]{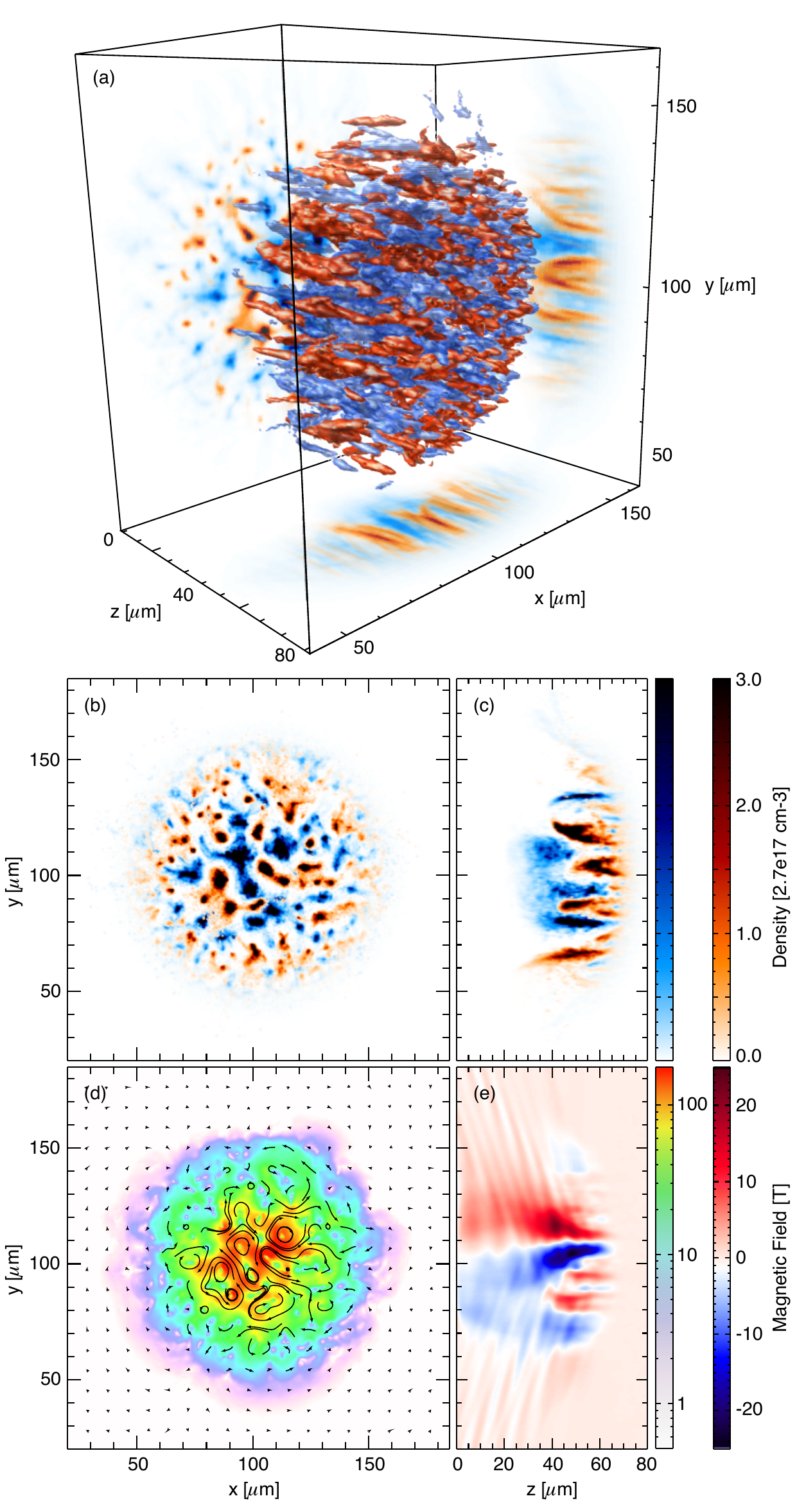}
\caption{Beam density and B-field after 10~cm propagation in a plasma with $n_e = 2.7\times10^{17}$~cm$^{-3}$. (a) Isosurfaces of $e^-$ (blue) and $e^+$ (red) density; projections correspond to the integration along the corresponding direction. (b-c) 2D central beam density slices ($e^-$ blue, $e^+$ red). (d) 2D central slice of radial B-field, $B_{\perp}=\sqrt{B_x^2+B_y^2}$, responsible for particle transverse motion and radiation (vectors represent B-field lines). (e) Integral of $B_y$ along $y$ ($ \int B_{y} dy/ \int dy$), measurable experimentally by Faraday rotation.} 
\label{fig:densityB}
\end{figure}

Fig.~\ref{fig:densityB}a-c show the structure of the {\em fireball beam} after 10~cm propagation in the laboratory plasma, or, equivalently, to the propagation of a fireball with a density 1~cm$^{-3}$ in $>50~$km in the background density of 1~cm$^{-3}$. The CFI generates well-defined current (and density) filaments, which size increases as the beam propagates in the plasma, and may grow to a  thickness above $5~\mu$m $\simeq 0.5~c/\omega_{\mathrm{pe}}$. 
These conditions correspond to a beam with $\sigma_r/(c/\omega_{\mathrm{pe}})\simeq 2$, which explains the few filaments obtained at saturation, reached when the filaments coalescence ceases and the B-field energy remains constant. The large currents associated with the beam filaments generate local B-fields up to 2~MGauss (Fig.~\ref{fig:densityB}d). The space charge separation, also associated with the filaments, leads to radial E-fields as high as $5 \times 10^8~\mathrm{V/cm}$. 
The presence of oblique modes/tilted filamentation \cite{bretsilva} is clear on Fig.~\ref{fig:densityB}c showing that, as expected, the beam can excite a combination of transverse (filamentation) and longitudinal (two-stream-like) instabilities.
 Finally, a system of filaments is present in the background plasma, behind the beam, evidencing a 3D structure. As the filaments merge, the space-charge separation leads to the plasma blowout and to the generation of strong E-fields.

This scenario is in stark contrast with that of a pure $e^-$ beam interacting with the same plasma, as in the recent PWFA experiments \cite{blumenfeld07}. Simulations and experiments for an $e^-$ beam with the same transverse size show that, for these parameters, the beam drives strong plasma wakefields that focus the beam to a narrow radius after one quarter betatron wavelength $\lambda_{\beta}/4=(2^{3/2}\pi\gamma^{1/2}c/\omega_{\mathrm{pe}})/4\simeq 5$~mm, and the beam envelope experiences oscillations along the plasma with period $\lambda_{\beta}/2$. No CFI is observed under these circumstances. 

\begin{figure}
\centering
\includegraphics[scale = 0.35]{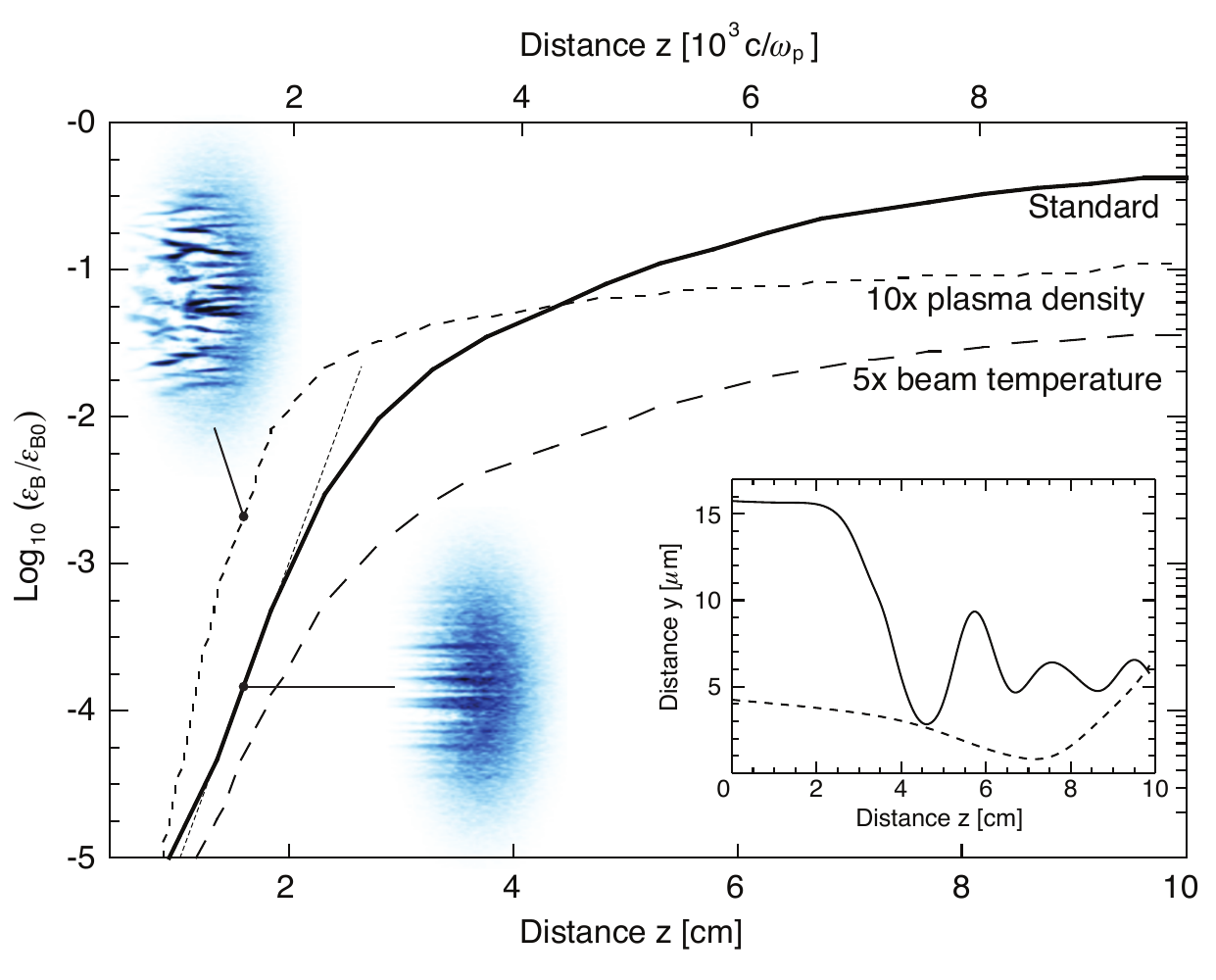}
\caption{Evolution of the equipartition parameter $\epsilon_B$, i.e., the total B-field energy ($B_x^2+B_y^2+B_z^2$) normalized to the kinetic energy of the particles $\epsilon_p = (\gamma_0-1)V_{b}$, ($V_{b}$ the volume of the beam) for different beam and plasma parameters. Values are normalized to $\epsilon_{B0}$, where $B_0$ is the field when the growth becomes exponential (after $\sim 0.1$~mm, or $10/\omega_{\mathrm{pe}}$). Standard case (solid line): {\em fireball beam} with   $2\times 10^{-5}$~m-rad emittance, in a plasma with $n_e = 2.7\times10^{17}$~cm$^{-3}$ (which also defines the baseline density for the normalization). The dotted line illustrates the linear growth rate. Slices of the density in the middle of the beam after $\sim1.5$~cm of plasma (plotted in blue) illustrate the difference in the instability structure. The inset includes the trajectories of two fireball electrons for the standard case.
}
\label{fig:parameter}
\end{figure}

In Fig.~\ref{fig:parameter}, we present the evolution with propagation distance in the plasma of the total normalized energy in the B-field, $\epsilon_B$ for different beam/plasma parameters, illustrating the exponential growth and saturation within the 10 cm range. 
The growth rates ($\Gamma_\mathrm{std}/\omega_{\mathrm{pe}}\simeq 2.0 \times 10^{-3}$, $\Gamma_\mathrm{hot}/\omega_{\mathrm{pe}}\simeq 1.7 \times 10^{-3}$, $\Gamma_\mathrm{high}/\omega_{\mathrm{pe}}\simeq 2.3 \times 10^{-3}$) are within the range predicted for this configuration for the purely transverse CFI 
($\Gamma_{\mathrm{max}}/\omega_{\mathrm{pe}} \simeq \sqrt 2 \beta_0/\sqrt{\gamma_0}[1+\beta_{\mathrm{th}}]$, with $\beta_{\mathrm{th}}=v_{\mathrm{th}}/c$ the particle thermal rms spread of velocity \cite{silva03}). A more detailed analysis reveals, however, that for higher plasma densities (keeping the beam density fixed) the growth rate is higher, but the saturated level of the B-field is lower. The former is an indication of the spatial-temporal character of the instability in this configuration, while the latter is an evidence for the different saturation mechanisms involved when the mixed mode/tilted filamention is dominant \cite{bretsilva}.

The finite transverse dimension of the beam determines (i) the longest wavenumber that can be excited, and (ii) the typical noise source for the instability. Since the beam is cold, the growth rate is already close to its maximum value for wavenumbers such that $k\le c/\omega_{pe}$. On the other hand, the finite length of the beam impacts the two-stream mode (or in the more general form of the filamentation instability, the oblique mode) \cite{Shukla2018}. However, there is no theory for the excitation of these modes for finite-length finite-width modes, and thus this work motivates further theoretical developments of a spatio-temporal theory for the {\em fireball beam} since it does not exist \cite{silva2009}.
The spatial-temporal theory for the two-stream instability \cite{jonesevans} predicts that an E-field perturbation excited at the vacuum/plasma interface ($x=0$ and $t=0$) will grow with $\propto \exp[i \omega_{\mathrm{pe}} \psi] \exp [(3 \sqrt{3}/4) (\psi x^2 n_e / n)^{1/3} \omega_{\mathrm{pe}}/\gamma ]$ where $\psi =t-z/v_b$ is the distance to the head of the beam. In the beam region, for the same distance in the laboratory and assuming the same initial perturbation in the longitudinal E-field, the amplified field in the standard case is approximately twice the amplified field in the high plasma density case. This is consistent with what we observe in the simulations, namely the fact that the mixed mode has clearly developed more strongly in the high density scenario. The coupling of the excited longitudinal field with the transverse field leads to the excitation of the mixed mode \cite{bretsilva}, as clearly seen in the high density case (Fig. \ref{fig:parameter}): the filaments are tilted, which indicates that the particles can detrap more easily, leading to a lower current and thus to a lower saturated B-field.


Our simulations assumed that the beams are aligned on-axis. Thus, the noise source for the CFI comes from the initial thermal fluctuations of the beam, which provide higher magnetic field seed values at larger wave numbers. An additional noise source for the instability will appear when the beams are not perfectly aligned. Using Ampere's law, it is straightforward to show that the corresponding noise source is stronger at wavelengths comparable to the beam width, $\sigma_r$. When the transverse size of the beam is much higher than the plasma skin depth, it is still possible to observe multiple filaments as the CFI growth rates are also smaller for smaller k. However, when $\sigma_r$ is comparable to the plasma skin depth, initial misalignments can quickly separate beam electrons from beam positrons, still leading to a two filamentary structure. In this case, to observe more filaments, ensuring that additional noise sources \cite{Allen} are present to ensure stronger initial magnetic field seeds at $k\geq 1/\sigma_r$ is required.

The interaction of the relativistic $e^-$ and $e^+$ with the B-fields confining the current filaments leads to the emission of synchrotron radiation. The oscillatory motion of the charges in the transverse directions due to the radial E-fields associated with the filaments of opposite charges leads to the emission of betatron radiation.
In both cases, the radiation is incoherent with the wiggler strength parameter $K \equiv a_\beta = \gamma k_\beta r \gg 1$, for $k_\beta$ the betatron wavenumber and $r$ the orbit amplitude. The spectra have a photon critical energy $E_{\mathrm{syn}} = \hbar \omega_{\mathrm{syn}} = \frac{3}{2} \hbar \gamma^2 |e| B/m_e c \simeq  (120~\mathrm{MeV}) (E[\mathrm{30GeV}])^2 B[\mathrm{2MG}]$ for the synchrotron radiation, and $E_{\beta} = \hbar \omega_{\beta}= \frac{3}{2} \hbar \gamma^3 r_\beta \simeq (0.6~\mathrm{MeV})  \eta \left(n_e[\mathrm{10^{17} cm^{-3}}]\right)^{1/2}$ for the betatron radiation, where $\eta \simeq 1$ describes the typical radius of the filaments (in units of $c/\omega_{\mathrm{pe}}$). Simulation results indicate that these two radiation processes might not be distinguishable \cite{jlmartins09}, at least for the initial stage, since $E_\perp$ and $B_\perp$ grow together. After a significant field growth in the CFI driven turbulence, however, the field structure may lead to different spectral signatures, as previously hinted in \cite{frederiksen,hededalthesis}. 

We briefly describe some of the particular aspects of the {\em fireball beam} diagnostic implementation. For PWFA applications the separation between the $e^-$ and the $e^+$ bunches must be of the order of a plasma wavelength ($\sim100~\mu$m). Such a small spacing between the bunches can be achieved with two interleaved magnetic chicanes with a coarse path length difference of the order of the bunch separation in the accelerator ($\sim5$~cm), and with fine magnetic adjustments \cite{decker}. These adjustments may also be used to overlap the two bunches in time and create the relativistic, neutral $e^-e^+$ {\em fireball beam}. The optimal temporal overlap is achieved by minimizing the coherent transition radiation the bunches emit when traversing a thin metallic foil located after the double chicane. The transverse overlap is obtained by imaging the incoherent optical transition radiation the bunches emit when traversing two thin foils located before and after the plasma. For PWFA experiments the beam ionizes a lithium vapor and creates the plasma by field-ionization \cite{blumenfeld07}, while the neutral {\em fireball beam} requires a pre-ionized plasma. Pre-ionization can be achieved by photo-ionization of a lithium vapor with an ultra-violet laser pulse \cite{muggli99}.

The filamentation of the beam is the most obvious indication of the CFI occurrence. The filaments, however, have a relatively small transverse size of $\sim5~\mu$m and, because of their emittance, diverge and overlap rapidly when exiting the plasma. To detect them inside the plasma, the strong plasma $e^-$ density gradients associated with the beam filamentation can be visualized with Schlieren shadowgraphy \cite{rienitz02} using a laser pulse propagating perpendicularly to the {\em fireball beam} path. The laser light is weakly deflected by the index of refraction variations corresponding to the $e^-$ density modulation.

The filamentation of the beam results in the generation of large B-fields in the plane perpendicular to the filaments themselves (see Fig.\ref{fig:densityB}d), which can be visualized by analyzing the polarization of a linearly polarized probe laser pulse traveling perpendicularly to the {\em fireball beam}, the same that is used for the Schlieren shadowgraphy. The laser light experiences Faraday rotation caused by the component of the filaments B-fields parallel to the laser propagation direction (see Fig.\ref{fig:densityB}e). Even though the B-field pattern is related to the structure of the random filaments, the effect computed from the simulation results for $B_y(y)$ produces an image similar to Fig.~\ref{fig:densityB}e and is clearly visible. Faraday rotation has been used to sample the B-fields generated in a laser wakefield experiment with similar parameters \cite{najmudin01}. The plasma density and B-field structure may be sampled along the $e^-$ beam path by moving the intersection point between the probe laser pulse and the plasma with a time resolution equal to the laser pulse length (fs) and a longitudinal resolution of the order of the probe beam size (mm), thereby giving access to the growth of the instability. 
The excess radiation associated with the oscillation of the $e^-$ and $e^+$ in the B-field and in the filaments can be directly observed using standard x-ray detection methods, similar to those that were used to detect synchrotron or betatron radiation in PWFA experiments \cite{Wang02}. Finally, the B-field growth occurs at the expense of beam energy. In our simulations the beam looses 6-11\% energy, i.e., 2-3~GeV, in the standard and high temperature cases, respectively. These beam energy changes can be measured using an imaging magnetic spectrometer, as in previous PWFA experiments \cite{Hogan05}.

In conclusion, we have shown that the $e^-e^+$ or {\em fireball beam} and plasma system that will be developed for PWFA experiments may also be used to produce in the laboratory a scenario relevant to test the very important microphysics issues of relativistic astrophysical phenomena. As a result of the CFI, the incoming {\em fireball beam} filamentation occurs over a plasma length of only a few cm. The current filaments generate large B-fields that lead to the enhanced emission of synchrotron and betatron radiation. Initial considerations indicate that the beam filamentation, the B-field generation, and the associated beam energy loss, as well as the excess radiation can in principle be observed in a single experiment.

We thank Prof. W. Mori and Dr. R. A. Fonseca for useful discussions.  Work supported by U.S. DoE Grant DE-FG02-92ER40745, by Funda\c c\~ao Calouste Gulbenkian, and by Funda\c c\~ao para a Ci\^encia e Tecnologia grants SFRH/BD/35749/2007 and PTDC/FIS/66823/2006 (Portugal). SFM and LOS thank KITP (UCSB) where part of this work was done, partially supported by NSF Grant PHY05-51164. Simulations performed on Dawson (UCLA) and IST (Portugal) clusters.


\end{document}